\documentclass[12pt]{revtex4}
\usepackage{amssymb}
\usepackage{latexsym}
\usepackage{epsfig}
\begin{document}                             

\title{Instability of QCD ghost dark energy model}

\author{Esmaeil Ebrahimi$^{1,2}$ \footnote{eebrahimi@uk.ac.ir} and Ahmad  Sheykhi$^{1,2,3}$ \footnote{
sheykhi@uk.ac.ir} }
\address{$^1$ Department of Physics, Shahid Bahonar University, PO Box 76175, Kerman, Iran\\
          $^2$ Research Institute for Astronomy and Astrophysics of Maragha (RIAAM), Maragha, Iran\\
          $^3$ Physics Department and Biruni Observatory, Shiraz University, Shiraz 71454,
          Iran}

\begin{abstract}
We investigate the instability of the ghost dark energy model
against perturbations in different cases. To this goal we use the
squared sound speed $v_s^2$ whose sign determines the stability of
the model. When $v_s^2<0$ the model is unstable against
perturbation. At first we discuss the noninteracting ghost dark
energy model in a flat FRW universe and find out that such a model
is unstable due to the negativity of the $v_s^2$ in all epoches. The
interacting ghost dark energy model in both flat and non-flat
universe are studied in the next parts and in both cases we find
that the squared sound speed of ghost dark energy is always
negative. This implies that the perfect fluid for ghost dark energy
is classically unstable against perturbations. In both flat and non
flat cases we find that the instability of the model increases with
increasing the value of the interacting coupling parameter.
\end{abstract}

 \maketitle
\section{Introduction}
Nowadays, there are enough observational evidences which indicate
that our universe is currently experiencing a phase of acceleration
\cite{Rie,cmb1,cmb2,sdss1,sdss2}. The unknown cause of this
unexpected acceleration called ``dark energy" (DE) whose nature is
still a matter of much doubt. Trying to explain observed
acceleration of the universe expansion two main approaches have been
followed in the literatures. The first is based on the DE proposal
which assumes that there exist an exotic form of energy with
negative pressure which is responsible for the acceleration of the
universe expansion. Famous alternatives in this approach including
cosmological constant $\Lambda$, scalar field models of DE such as
quintessence \cite{wetter,ratra}, K-essence
\cite{chiba,armend1,armend2}, phantom fields \cite{cald}, tachyon
\cite{tachyon}, holographic DE  \cite {HDE} and agegraphic DE
\cite{ADE} and so on (for a recent review on DE models see
\cite{nederev} and references therein). The second approach for
explanation of the acceleration expansion is based on the
modification of the gravity theory. In this approach the DE problem
is considered as a shortcoming of the Einstein's gravity and try to
modify the standard model of cosmology in such a way that the phase
of acceleration is reproduced without including any new kind of
energy. Examples of these models are $f(R)$ gravity
\cite{capoz,carroll} and braneworld scenarios \cite{DGP,sheywang}.

Seeking a solution to $U(1)$ problem , the so-called Veneziano ghost
has been proposed in the low energy effective QCD where they are
completely decoupled from the physical sector
\cite{kawar,witten,rosen,nath}. The ghosts make no contribution in
the flat Minkowski space, but make a small energy density
contribution to the vacuum energy due to the off-set of the
cancelation of their contribution in curved space or time-dependent
background. This contribution to the vacuum energy density can be
considered a possible candidate for the origin of the cosmological
constant \cite{Ohta}. In a dynamic background or a spacetime with
non-trivial topology the ghost field contributes to the vacuum
energy proportional to $\Lambda^3_{QCD} H$, where $H$ is the Hubble
parameter and $\Lambda^3_{QCD}$ is $QCD$ mass scale. With
$\Lambda_{\rm QCD}\sim 100 MeV$ and $H \sim 10^{-33}eV$ ,
$\Lambda^3_{\rm QCD}H$ gives the right order of magnitude $\sim
(3\times10^{-3}eV)^4$ for the observed DE density \cite{Ohta}. This
remarkable coincidence implies that the ghost dark energy (GDE)
model is free from the fine tuning problem \cite{Urban,Ohta}. It was
shown that this vacuum energy density can play the role of DE in the
evolution of the universe \cite{CaiGhost,sheykhigde}.

Every new model of DE represents new features and consequences which
should be explored carefully. One way to test the viability of a new
DE model is to explore its stability against perturbations. To
investigate the stability, a key quantity is the squared speed of
sound $v_s^2 = dp/d\rho$ \cite{peebleratra}. The sign of $v_s^2$
plays a crucial role in determining the stability of the background
evolution. If $v_s^2<0$, it means the classical instability of a
given perturbation. This issue has already  been investigated for
some DE models. It was shown that chaplygin gas
 and tachyon DE have positive squared speeds of sound with, $v_s^2=
-w$, and thus they are supposed to be stable against small
perturbations \cite{gorini,sandvik}. However, the perfect fluid of
holographic DE with future event horizon is classically unstable
because its squared speed is always negative \cite{myung1} . Also in
\cite{myung2}, it is shown that the agegraphic model of DE have a
negative sound speed squared in flat, non-flat and also in the
presence of interaction indicating the instability of this model
against perturbations.

DE and dark matter (DM) are usually considered as two distinct dark
components of the universe due to their unknown nature and very
different origin of them. Note that DM has gravity effect and plays
a crucial role in explanation of the galaxy rotation curve, while DE
has anti-gravity feature which pushes the universe and makes it
expansion accelerated. However, in recent years several signals have
been detected, implying a small interaction between DE and DM is
possible. As an instance, observational evidences provided by the
galaxy cluster Abell A586 supports the interaction between DE and DM
\cite{interact1}. Beside that, lately there arose an interest in the
non-flat version models of DE in literature. This new enthusiasm
originate from some observations which challenge the idea of the
flat universe. For example evidences from CMB and also supernova
measurements of the cubic correction to the luminosity distance
favor a positively curved universe \cite{nonflat2,nonflat3}. In
addition, some exact analysis of the WMAP data reveals the
possibility of a closed universe \cite{nonflat4}.

All above reasons motivate us to study the stability of interacting
GDE model in a nonflat universe. In this paper, we would like to
generalize the approach presented in
\cite{peebleratra,myung1,myung2} to GDE model in a universe with
spacial curvature in the presence of interaction between the dark
matter and DE. Various aspects of GDE have recently investigated. A
thermodynamical description of GDE is discussed in \cite{feng}.
Tachyon and quintessence reconstruction of GDE model were studied in
\cite{tachgde} and \cite{quintgde} respectively. The study has also
been extended to Brans-Dicke theory \cite{ebrahimibdgde}.

This paper is organized as follows. In the next section, we review
the GDE model in both flat and nonflat universe. In section
\ref{sta}, we explore the stability of the GDE model in all
discussed cases of section II. We summarize our results in section
\ref{sum}.
\section{A brief review on ghost dark energy}\label{rev}

\subsection{Noninteracting ghost dark energy}
\begin{figure}\epsfysize=5cm
{ \epsfbox{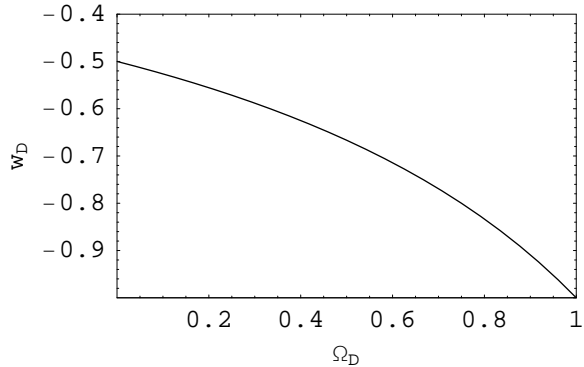}}\caption{Evolution of $w_D$ versus $\Omega_D$
for noninteracting GDE.} \label{f1}
\end{figure}

\begin{figure}\epsfysize=5cm
{ \epsfbox{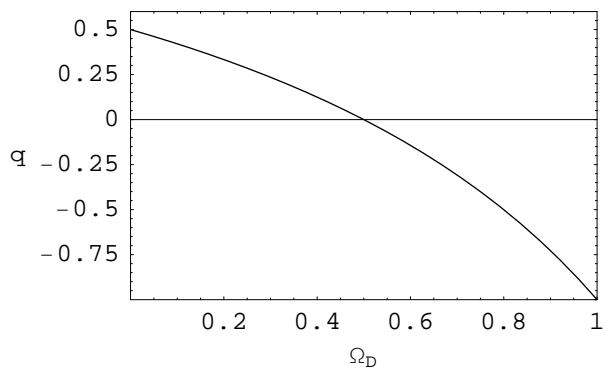}}\caption{Evolution of deceleration parameter
against $\Omega_D$ for noninteracting GDE.} \label{f2}
\end{figure}

Let us at first review the noninteracting GDE in a flat FRW
background filled with a matter component and GDE. We follow the
method of \cite{sheykhigde}. Dynamic of such a universe is
determined by the Friedmann equation
\begin{equation}\label{frideqs}
H^2=\frac{1}{3M_p^2} \left( \rho_m+\rho_D \right),
\end{equation}
where $\rho_m$ is the energy density of pressureless DM and
$\rho_D$ is the GDE density. According to standard cosmology one
can define the fractional density of different energy components
of the universe as
\begin{equation}\label{Omega}
\Omega_m=\frac{\rho_m}{\rho_{cr}},\ \ \
\Omega_D=\frac{\rho_D}{\rho_{cr}},
\end{equation}
where the critical energy density is $\rho_{cr}={3H^2 M_p^2}$.
Thus, the Friedmann equation can be rewritten as
\begin{equation}\label{fridomega}
\Omega_m+\Omega_D=1.
\end{equation}
Here, we consider the GDE which its energy density can be written as
\cite{Ohta}
\begin{equation}\label{gde}
    \rho_D=\alpha H,
\end{equation}
where $\alpha$ is a constant of order $\Lambda_{\rm QCD}^3$ and
$\Lambda_{\rm QCD}$ is QCD mass scale. Based on the definition of
$\Omega_D$ and using (\ref{gde}), we get
\begin{equation}\label{omegad}
    \Omega_D=\frac{\rho_D}{\rho_{cr}}=\frac{\alpha}{3M_p^2 H}.
\end{equation}
Energy conservation equations for different components read
\begin{eqnarray}
\dot\rho_m+3H\rho_m&=&0,\label{consm}\\
\dot\rho_D+3H\rho_D(1+w_D)&=&0\label{consd}.
\end{eqnarray}
Taking the time derivative of relation (\ref{gde}), and using the
Friedmann equation (\ref{frideqs}) as well as the continuity
equation (\ref{consd}) we find
\begin{equation}\label{eq1}
  \frac{\dot{\rho}_D}{\rho_D}=\frac{\dot{H}}{H}=-\frac{3}{2}H\left[1+\Omega_D
  w_D\right].
\end{equation}
Substituting this relation in continuity equation (\ref{consd}) we
obtain the equation of state (EoS) parameter of GDE, namely
\begin{equation}\label{wDni}
w_D=-\frac{1}{2-\Omega_D}.
\end{equation}
Evolution of $w_D$ versus $\Omega_D$ is shown in Fig.\ref{f1}. One
finds that at the early time  where $\Omega_D\ll 1$ we have
$w_D=-1/2$, while at the late time where $\Omega_D\rightarrow 1$ the
GDE mimics a cosmological constant, namely $w_D= -1$.
Another interesting quantity to be calculated is the deceleration
parameter
\begin{equation}\label{q1}
q=-1-\frac{\dot{H}}{H^2}=\frac{1}{2}-\frac{3}{2}\frac{\Omega_D}{(2-\Omega_D)}.
\end{equation}
From above relation it is clear that at the late time where
$\Omega_D\rightarrow1$, $q=-1$. At the early universe where
$\Omega_D\rightarrow0$ we have $q={1}/{2}$, which corresponds to the
matter dominated epoch. Evolution of $q$ versus $\Omega_D$ is
plotted in Fig.\ref{f2}.
Two important points should be emphasized for noninteracting GDE
model. First, in this model there exist no free parameter. Second,
$w_D$ cannot  cross the phantom line. The evolution of GDE is
governed by \cite{sheykhigde}
\begin{equation}\label{OmegaDev}
\Omega^{\prime}_D=3\Omega_D\left(\frac{1-\Omega_D}{2-\Omega_D}
\right),
\end{equation}
where the prime denotes the derivative with respect to $x=\ln a$.
\subsection{Interacting ghost dark energy}
DM and DE  are usually considered separately in the modern
cosmology. However, there exist some observational evidences which
support an interaction between them \cite{berto1}.Thus, it seems
meaningful to study the interacting version of GDE. To this goal we
start with energy conservation equations

\begin{eqnarray}
&&\dot{\rho}_m+3H\rho_m=Q, \label{consmint}
\\&& \dot{\rho}_D+3H\rho_D(1+w_D)=-Q.\label{consqint}
\end{eqnarray}
In these equations the form of $Q$ has not been chosen exactly yet,
however, we know that $Q$ should be small and positive. A large and
positive choice of $Q$ does not lead to late time acceleration while
a large and negative choice of $Q$ will lead to an early domination
of DE component which does not let the large structures to be formed
in the universe. One should note that a choice of negative sign for
$Q$ implies decaying of DE to the cold DM.  In this paper we choose
the interaction term as $Q=3H^2(\rho_D+\rho_M)$ which can be written
as $Q=3H^2\rho_D(1+\frac{\Omega_M}{\Omega_D})$. In the case of
interacting, the EoS parameter is obtained as \cite{sheykhigde}
\begin{equation}\label{wDi}
w_D=-\frac{1}{2-\Omega_D}\left(1+\frac{2b^2}{\Omega_D}\right).
\end{equation}
Setting $b=0$, $w_D$ for the noninteracting case is retrieved. In
the late time where $\Omega_D\rightarrow 1$, the EoS parameter of
interacting GDE necessary crosses the phantom line, namely,
$w_D=-(1+2b^2)<-1$ independent of the value of coupling constant
$b^2$. Also the deceleration parameter can be calculated as
\cite{sheykhigde}
\begin{equation}\label{qint}
q=\frac{1}{2}-\frac{3}{2}\frac{\Omega_D}{(2-\Omega_D)}\left(1+\frac{2b^2}{\Omega_D}\right).
\end{equation}
The equation of motion of $\Omega_D$ can be obtained as
\begin{equation}\label{omegadevin}
\frac{d\Omega_D}{d\ln
a}=\frac{3}{2}\Omega_D\left[1-\frac{\Omega_D}{(2-\Omega_D)}\left(1+\frac{2b^2}{\Omega_D}\right)\right].
\end{equation}
\subsection{Interacting ghost dark energy in non-flat universe}
In the past decade several observational evidences have been
observed in contrast to the flatness assumption of the universe. In
the context of inflation which assumes a large rate of expansion at
the beginning instants of the evolution of the universe, it is
argued that the flatness is not a necessary consequence of inflation
if the number of e-folding is not very large \cite{huang}. Besides,
the parameter $\Omega_k$ represents the contribution to the total
energy density from the spatial curvature and it is constrained as
$-0.0175 <\Omega_k< 0.0085$ with $95\%$ confidence level by current
observations \cite{water}. These motivate us to study the GDE model
in the presence of curvature. Taking the curvature into account, the
first Friedmann equation is written as
\begin{eqnarray}\label{Friedm}
H^2+\frac{k}{a^2}=\frac{1}{3M_P^2} \left( \rho_m+\rho_D \right).
\end{eqnarray}
This equation can be written as
\begin{equation}\label{fridomeganf}
\Omega_m+\Omega_D=1+\Omega_k,
\end{equation}
where $\Omega_k=\frac{k}{a^2H^2}$. We can obtain different
parameters in the presence of the curvature term. Taking the time
derivative of the Friedmann equation (\ref{Friedm}) and using
(\ref{fridomeganf}), we find
\begin{equation}\label{doth2}
\frac{\dot{H}}{H^2}=\Omega_k-\frac{3}{2}
\left[1+\Omega_k+\Omega_Dw_D\right].
\end{equation}
Combining the above equation with Eq. (\ref{consqint}), after
using $Q=3H^2\rho_D(1+\frac{\Omega_M}{\Omega_D})$, as well as
$\frac{\dot{\rho}}{\rho}=\frac{\dot{H}}{H}$ we obtain the EoS
parameter as
\begin{equation}\label{wDnfint}
w_D=-\frac{1}{2-\Omega_D}\left(1-\frac{\Omega_k}{3}+\frac{2b^2}{\Omega_D}
(1+\Omega_k)\right).
\end{equation}
The deceleration parameter can be calculated by substituting Eqs.
(\ref{doth2}) and (\ref{wDnfint}). We find
\begin{equation}\label{q2}
q=\frac{1+\Omega_k}{2}+\frac{3\Omega_D}{2(2-\Omega_D)}\left[1-\frac{\Omega_k}{3}+\frac{2b^2}{\Omega_D}(1+\Omega_k)\right].
\end{equation}
Also the equation of motion of GDE can be obtained as
\cite{sheykhigde}
\begin{equation}\label{Omegaprime2n}
\frac{d\Omega_D}{d\ln
a}=\frac{3}{2}\Omega_D\left(1+\frac{\Omega_k}{3}-\frac{\Omega_D}{2-\Omega_D}\left[1-\frac{\Omega_k}{3}+2b^2
\Omega_D^{-1} \left(1+\Omega_k\right)\right]\right).
\end{equation}
It is worth mentioning that all relations we obtained in this
subsection restor their respective expressions in the previous
subsection when we set $\Omega_k=0$.
\section{Instability of the ghost dark energy}\label{sta}
The main idea for investigating the stability of GDE model comes
from the perturbation theory. Assuming a small perturbation in the
background energy density, we would like to see if the perturbation
grows or will collapse. In the linear perturbation theory, the
perturbed energy density of the background can be written as
\begin{equation}\label{pert1}
    \rho(t,x)=\rho(t)+\delta\rho(t,x),
\end{equation}
where $\rho(t)$ is unperturbed background energy density. The energy
conservation equation ($\nabla_{\mu}T^{\mu\nu}=0$) yields
\cite{peebleratra}
\begin{equation}\label{pert2}
\delta\ddot{\rho}=v_s^2\nabla^2\delta\rho(t,x),
\end{equation}
where $v_s^2=\frac{dP}{d\rho}$ is the squared of the sound speed.
Solutions of equation (\ref{pert2}) include two cases of interest.
First when $v_s^2$ is positive  Eq.
(\ref{pert2}) becomes an
ordinary wave equation whose its solutions would be oscillatory
waves of the form $\delta \rho=\delta\rho_0e^{-i\omega
t+i\vec{k}.\vec{x}}$ which indicates a propagation mode for the
density perturbations. The second is when $v_s^2$ is negative. In
this case the frequency of the oscillations becomes pure imaginary
and the density perturbations will grow with time as $\delta
\rho=\delta\rho_0e^{\omega t+i\vec{k}.\vec{x}}$. Thus the growing
perturbation with time indicates a possible emergency of
instabilities in the background.

In what follows, we are trying to see if the GDE leads to
instabilities in the FRW background. To this end we will calculate
the sound speed for both interacting and noninteracting GDE model.
\subsection{Instability of noninteracting ghost dark energy}
In this section we would like to obtain the sound speed in a
background filled with matter and GDE in the absence of interaction.
The squared sound speed can be written as

\begin{equation}\label{speeddef}
    v_s^2=\frac{dP}{d\rho}.
\end{equation}

In order to obtain the sound speed in a background filled with
barotropic fluids we rewrite the definition as \cite{CaiGhost}
\begin{equation}\label{speeddef2}
    v_s^2=\frac{dP}{d\rho}=\frac{\dot{P}}{\dot{\rho}}=\frac{\rho}{\dot{\rho}}\dot{w}+w,
\end{equation}
where in the last step we have used $P=w\rho$. Taking time
derivative of Eq.({\ref{gde}}) and using (\ref{consd}), we obtain
\begin{equation}\label{hdot}
    \frac{H}{\dot{H}}=\frac{\rho}{\dot{\rho}}=\frac{-2}{3H(1+\Omega_D w_D)}
\end{equation}
\begin{figure}\epsfysize=5cm
{ \epsfbox{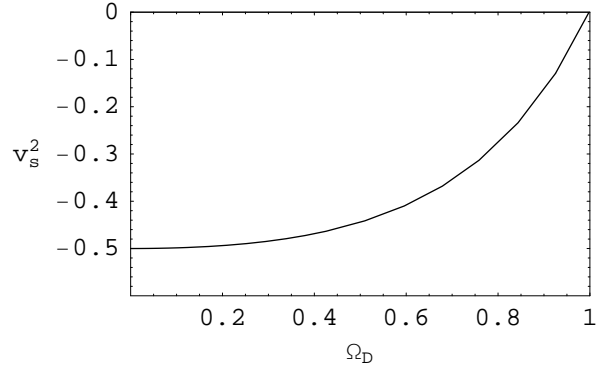}}\caption{Evolution of squared sound speed
$v_s^2$ against $\Omega_D$ for noninteracting GDE model.}
\label{fvnin}
\end{figure}
Inserting (\ref{wDni}), into above relation we get
\begin{equation}\label{rrhodot}
\frac{\rho}{\dot{\rho}}=-\frac{2-\Omega_D}{3H(1-\Omega_D)}
\end{equation}
Taking the time derivative of (\ref{wDni}) yields
\begin{equation}\label{wdot}
    \dot{w}_D=-\frac{\dot{\Omega}_D}{(2-\Omega_D)^2}.
\end{equation}
Replacing Eqs.(\ref{rrhodot}) and (\ref{wdot}) into
(\ref{speeddef2}) and also using (\ref{OmegaDev}) we get
\begin{equation}\label{v2ni}
    v_s^2=-2\frac{1-\Omega_D}{(2-\Omega_D)^2},
\end{equation}
where we also used $\frac{d}{dt}=H\frac{d}{d\ln a}$. This result is
the same as one presented in \cite{CaiGhost}. Having the $v_s^2$ at
hand we are ready to discuss about the stability of perturbations.
It is clear that in the evolution history of the universe $\Omega_D$
can achieve values in the range $[0,1]$ which $\Omega_D=0$ indicates
beginning stages of the evolution of the universe while $\Omega_D=1$
is an extrapolation of the future of the universe. One can easily
see from (\ref{v2ni}) that $v_s^2$ is always negative and varies
between $[-\frac{1}{2},0]$. Hence, Eq. (\ref{pert2}) has solution of
the second category ($\delta \rho=\delta\rho_0e^{\omega
t+i\vec{k}.\vec{x}}$). This result indicates that due to the
negativity of the squared sound speed every small perturbation can
grow with time which leads to an instability in the universe. Thus
we cannot expect a noninteracting GDE dominated universe in the
future as the fate of the universe. The evolution of the $v_s^2$
versus $\Omega_D$ is shown in Fig. \ref{fvnin}.
\subsection{Instability of interacting ghost dark energy}
A same steps as the pervious section can be followed to obtain the
squared sound speed $v_s^2$ for the interacting case. Taking time
derivative of Eq. (\ref{wDi}) we have
\begin{equation}\label{wdoti}
    \dot{w}_D=
    -\frac{\dot{\Omega}_D}{(2-\Omega_D)^2}\left[1+\frac{4b^2}{\Omega_D^2}(\Omega_D-1)\right].
\end{equation}
Also from (\ref{eq1}) one finds
\begin{equation}\label{rrhodot2}
    \frac{\rho}{\dot{\rho}}=-\frac{2}{3H(1+\Omega_Dw_D)}.
\end{equation}
Taking into account relation $\frac{d}{dt}=H\frac{d}{d\ln a}$ as
well as (\ref{omegadevin}) yields
\begin{equation}\label{omegadotin}
\dot{\Omega}_D=\frac{3}{2}H\Omega_D\left[1-\frac{\Omega_D}{(2-\Omega_D)}\left(1+\frac{2b^2}{\Omega_D}\right)\right].
\end{equation}
Replacing these relations in (\ref{speeddef2}) and after a little
algebra one obtains
\begin{equation}\label{v2in}
    v_s^2=-2\frac{1-\Omega_D}{(2-\Omega_D)^2}+2b^2\frac{3\Omega_D-4}{\Omega_D(2-\Omega_D)^2},
\end{equation}
\begin{figure}\epsfysize=5cm
{ \epsfbox{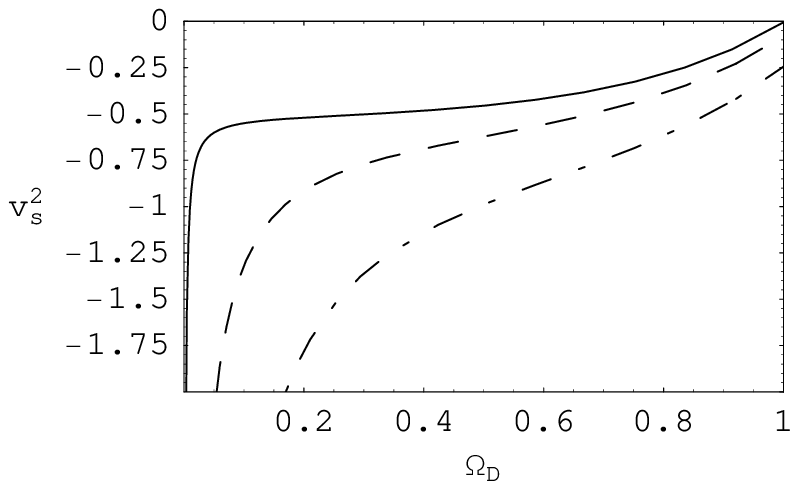}}\caption{This figure shows evolution of squared
sound speed $v_s^2$ versus $\Omega_D$ for interacting GDE model. The
solid line corresponds to $b=0.05$, dashed line to $b=0.2$ and
dashed-dot line to $b=0.35$.} \label{fvin}
\end{figure}
which is the squared sound speed for interacting GDE fluid. The
evolution of $v_s^2$ against $\Omega_D$ is plotted in Fig.
(\ref{fvin}) for different values of the coupling parameter $b$. The
figure reveals that $v_s^2$ is always negative and thus, as the
previous case, a background filled with the interacting GDE seems to
be unstable against the perturbation. This implies that we cannot
obtain a stable GDE dominated universe. One important point is the
sensitivity of the instability to the coupling parameter $b$. The
larger $b$, leads to more instability against perturbations .
\subsection{Instability of interacting ghost dark energy in non-flat universe}
Finally, we study the instability of interacting GDE model in a
universe with spacial curvature. From Eq. (\ref{doth2}) we can
obtain
\begin{equation}\label{rrdotnf}
\frac{\rho}{\dot{\rho}}=\frac{-2}{3H\left[1+\frac{\Omega_k}{3}-\frac{\Omega_D}{2-\Omega_D}
\left(1-\frac{\Omega_k}{3}+\frac{2b^2}{\Omega_D}(1+\Omega_k)\right)\right]}
\end{equation}
Taking the time derivative of Eq. (\ref{wDnfint}), yields
\begin{equation}\label{wdotnfint}
    \dot{w}_D=\frac{\dot{\Omega}_D}{2-\Omega_D}\left[-\frac{1}{2-\Omega_D}\left(1-\frac{\Omega_k}{3}+\frac{2b^2}{\Omega_D}
(1+\Omega_k)\right)+\frac{2b^2}{\Omega_D^2} (1+\Omega_k)\right]
\end{equation}
Having the above relations at hand we are in a position to obtain
the squared sound speed $v_s^2$. Replacing Eqs. (\ref{rrdotnf}),
(\ref{wdotnfint}) in (\ref{speeddef2}), after using
(\ref{Omegaprime2n}), one gets
\begin{equation}\label{v2nfint}
    v_s^2=-\frac{2(1-\Omega_D)}{(2-\Omega_D)^2}+\frac{2}{3}\frac{1-\Omega_D}{(2-\Omega_D)^2}\Omega_k
    +\frac{2b^2}{\Omega_D}\frac{3\Omega_D-4}{(2-\Omega_D)^2}(1+\Omega_k).
\end{equation}
Setting $\Omega_k=0=b$ the above relation reduces to the flat
noninteracting respective relation. Also the squared sound speed of
the flat interacting case can be retrieved when $\Omega_k=0$.

\begin{figure}\epsfysize=5cm
{ \epsfbox{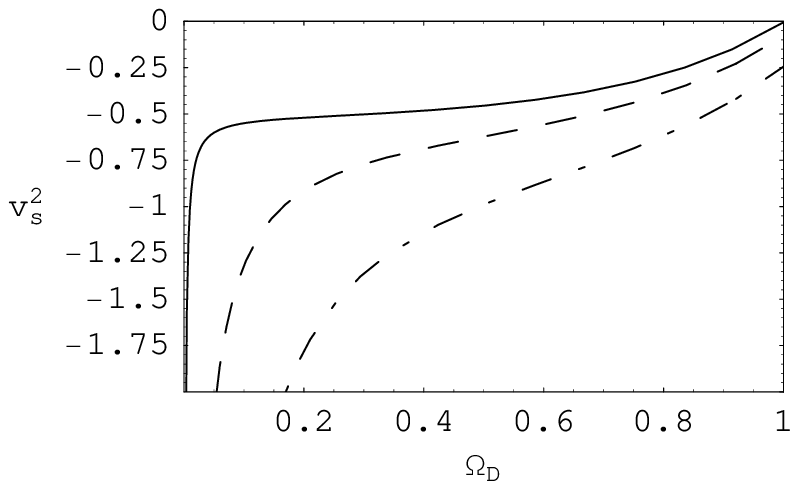}}\caption{This figure shows evolution of squared
sound speed $v_s^2$ versus $\Omega_D$ for interacting GDE model in a
nonflat background. The solid line corresponds to $b=0.05$, dashed
to $b=0.2$ and dashed-dot line to $b=0.35$.} \label{fvnfin}
\end{figure}

In order to obtain an insight on the stability issue of the
interacting GDE in a nonflat FRW universe, we have to discuss on the
sign of $v_s^2$ during the evolution of the universe. To this end we
plot $v_s^2$ versus $\Omega_D$, where the value of $\Omega_D$
indicates different epoches of evolution. The result can be seen in
Fig. \ref{fvnfin} which clearly indicates an almost same behavior as
the flat interacting case. For all epoches the squared sound speed,
($v_s^2$), is negative indicating the instability of the universe
against perturbations in the GDE background. Once again we see the
crucial dependence of the stability to the coupling parameter $b$.
Increasing $b$ will result more instability in the universe which is
clearly seen in Fig. \ref{fvnfin}. As a result the universe filled
with DM and GDE even in the presence of the curvature cannot lead to
a stable GDE dominated universe.
\section{Summary and Discussion} \label{sum}
Every new DE model should at first explain the acceleration of the
universe expansion whose EoS parameter satisfies $w_D<-1/3$. If the
model passed this test, then it should be investigated for further
features and consequences. Since we know from observation that our
stable universe is experiencing a phase of acceleration due to the
domination of DE (assuming the DE approach for explanation of the
acceleration of the universe is correct), thus every new DE model
should be capable to lead a stable DE dominated universe. Thus the
investigation on the stability of a proposed DE model is well
motivated. Among various DE model the so called QCD ghost DE model
was recently proposed to explain the DE dominated universe within
the framework of standard model of particle physics and general
relativity. It was argued that the vacuum energy of the Veneziano
QCD ghost field in a time-dependent background can play the role of
DE with an energy density proportional to the Hubble parameter $H$
\cite{Ohta}. The advantages of this new proposal compared to the
previous DE models is that it is totally embedded in standard model
so that one needs not to introduce any new parameter, new degree of
freedom or to modify general relativity \cite{CaiGhost}.

In this paper we have explored the stability of the GDE model
against perturbations in the flat/nonflat background and
with/without interaction. We used the squared sound speed
($v_s^2=\frac{dP}{d\rho}$) as the main factor for studying the
stability. If $v_s^2$ is positive the GDE would be stable against
perturbations. When $v_s^2$ is negative we encounter the instability
in the background spacetime. We have discussed several cases
including whether there is or not an interaction between DM and GDE
and whether there is or not a curvature term in the background
metric. Interestingly enough, we found that the GDE model is always
unstable against perturbations. As a result, the universe filled
with DM and GDE component cannot lead to a stable GDE dominated
universe. We also observed that the instability of the interacting
GDE increases with increasing the interacting coupling parameter
$b$.
\acknowledgments{ This work has been supported by Research
Institute for Astronomy and Astrophysics of Maragha, Iran.}

\end{document}